\documentclass[10pt,twocolumn]{article}

\usepackage{xcolor}
\usepackage{amsmath}
\usepackage{graphicx}
\usepackage[margin=0.7in]{geometry}
\usepackage[numbers]{natbib}

\begin{document}

\title{Natural brain-information interfaces: Recommending information by
  relevance inferred from human brain signals}

\author{
  Manuel J. A. Eugster$^{1}$\footnote{equal
    contribution} \footnote{corresponding authors},
  Tuukka Ruotsalo$^{1}$\footnotemark[1],
  Michiel M. Spap\'{e}$^{1}$\footnotemark[1],\\
  Oswald Barral$^2$,
  Niklas Ravaja$^{1,3}$,
  Giulio Jacucci$^{1,2}$,
  Samuel Kaski$^{1,2}$\footnotemark[2]
}

\date{\small $^1$~Helsinki Institute for Information
    Technology HIIT, Department of Computer Science, Aalto University,
    Finland. $^2$~Helsinki Institute for Information Technology
    HIIT, Department of Computer Science, University of Helsinki,
    Finland. $^3$~Helsinki Institute for Information
     Technology HIIT, Department of Social Research, University of
     Helsinki, Finland.}

\maketitle

\begin{abstract}
  Finding relevant information from large document
  collections such as the World Wide Web is a common task in our daily
  lives. Estimation of a user's interest or search intention is
  necessary to recommend and retrieve relevant information from these
  collections. We introduce a brain-information interface used for
  recommending information by relevance inferred directly from brain
  signals.  In experiments, participants were asked to read Wikipedia
  documents about a selection of topics while their EEG was recorded.
  Based on the prediction of word relevance, the individual's search
  intent was modeled and successfully used for retrieving new,
  relevant documents from the whole English Wikipedia corpus.  The
  results show that the users' interests towards digital content can
  be modeled from the brain signals evoked by reading. The introduced
  brain-relevance paradigm enables the recommendation of information
  without any explicit user interaction, and may be applied across
  diverse information-intensive applications.
\end{abstract}

\section{Introduction}

Documents on the World Wide Web, and seemingly countless other
information sources available in a variety of on-line services, have
become a central resource in our day-to-day decisions. As our
capabilities are limited in finding relevant information from large
collections, computational recommender systems have been introduced to
alleviate information
overload~\cite{Hanani:2001:IFO:598287.598363}. To predict our future
needs and intentions, recommender systems rely on the history of
observations about our
interests~\cite{Schein:2002:MMC:564376.564421}. Unfortunately, people
are reluctant to provide explicit feedback to recommender
systems~\cite{Kelly:2003:IFI:959258.959260}. As a consequence,
acquiring information about user intents has become a major bottleneck
to recommendation performance, and sources of information about the
individual's interests have been limited to the implicit monitoring of
online behavior, such as which documents they read, which videos they
watch, or for which items they
shop~\cite{Kelly:2003:IFI:959258.959260}.  An intriguing alternative
is to monitor the brain activity of an individual; that could mitigate the
cognitive load involved in expressing intentions and enable the direct
inferring of information about relevance.

\begin{figure}
  \centerline{\includegraphics[width=0.48\textwidth]{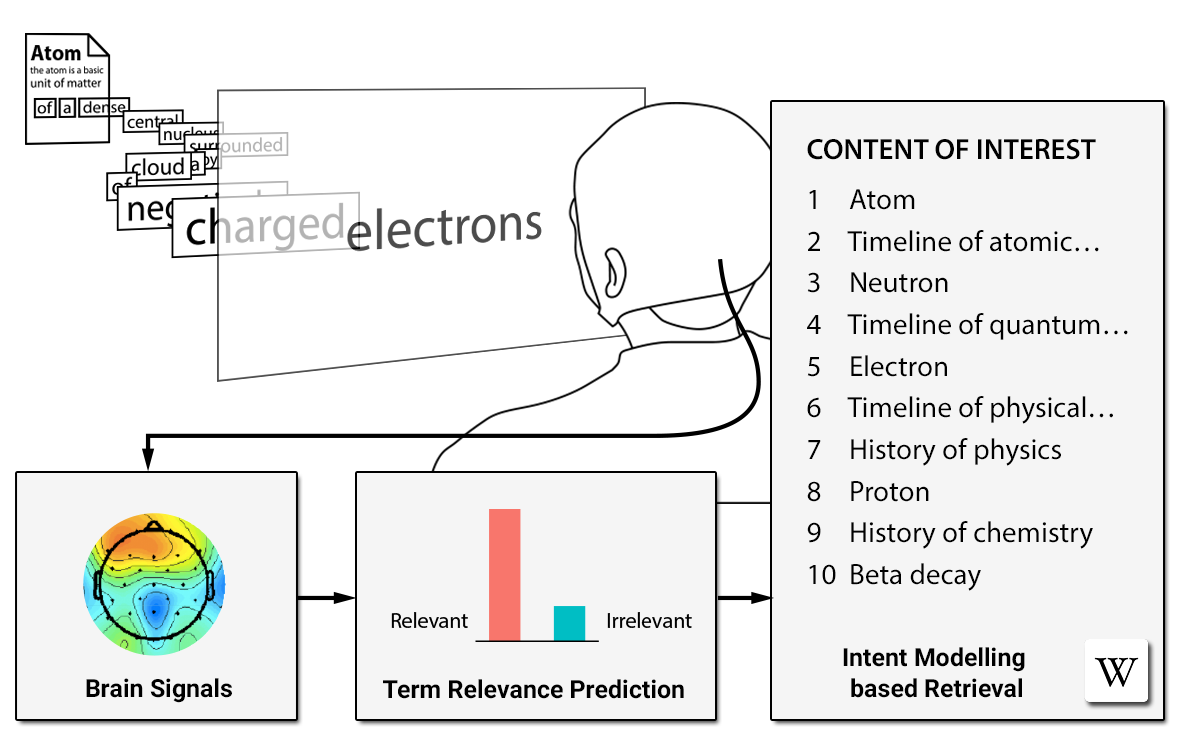}}
  \caption{The user reads text from the English Wikipedia while the
    event-related potentials (ERPs) are recorded using
    electroencephalography (EEG). A
    classifier is trained to distinguish the relevant from the
    irrelevant words by using the ERPs associated with each word in
    the text. An intent model uses the
    relevance estimates as input and then infers the user's search intent. The
    intent model is used to retrieve new information from the English
    Wikipedia.\label{fig:mindir-illu}}
\end{figure}

To utilize brain signals, we introduce a brain-relevance
paradigm for information filtering. The paradigm is based on the
hypothesis
that relevance feedback on individual words, estimated from brain
activity during a reading task, can be utilized to automatically
recommend a set of documents relevant to the user's topic of
interest (see Figure~\ref{fig:mindir-illu} for an illustration).
Following the brain-relevance paradigm, we introduce the first
end-to-end methodology for performing fully automatic information
filtering by using only the associated brain 
activity (Figure~\ref{fig:mindir-illu}). The methodology is based on
predicting and modeling the user's 
informational intents~\cite{Ruotsalo+:2015} using brain signals and
the associated text corpus statistics, and recommending new and unseen 
information using the estimated intent model.

We demonstrate the
effectiveness of the methodology with brain signals naturally evoked
during a text reading task. That is, unlike standard active
brain-computer interface (BCI)
practices, the method used here does not require the user to perform
additional, explicit tasks (such as the mental counting of relevant words) that
have been previously shown to enhance the signal-to-noise
ratio~\cite{Zander+:2010}. Instead, the methodology relies solely on
the detection of the neural activity patterns associated with relevance,
so that applications benefit from truly implicit and passive
measurements.

The data from experiments, in which electroencephalography (EEG) was
recorded from 15 
participants while they were reading texts, shows that the recommendation
of new relevant information can be significantly improved using brain
signals when compared to a random baseline. The result suggests that
relevance can be predicted from brain signals that are naturally
evoked when users read, and they can be utilized in recommending new information
from the Web as a part of our everyday information-seeking activities.

\begin{figure*}
  \centerline{
\raisebox{5cm}{\textbf{a}}
\includegraphics[scale=0.34]{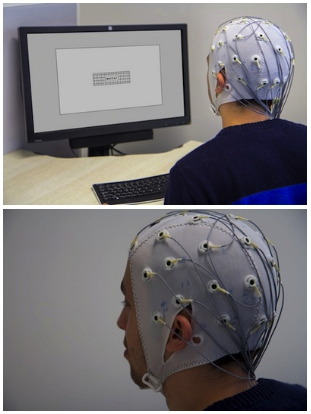}
\raisebox{5cm}{\textbf{b}}
\includegraphics[scale=0.6]{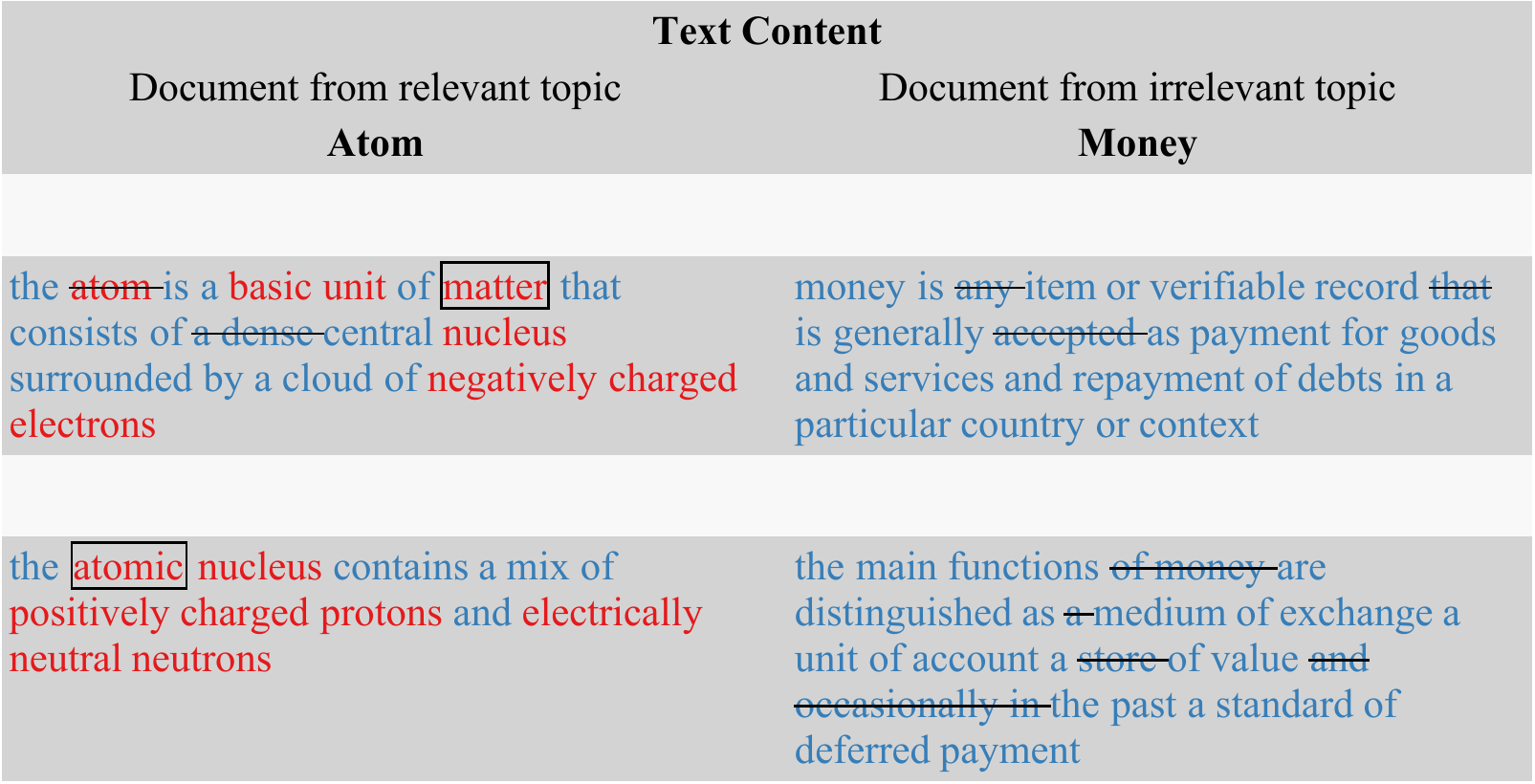}
\raisebox{5cm}{\textbf{c}}
\includegraphics[scale=0.57]{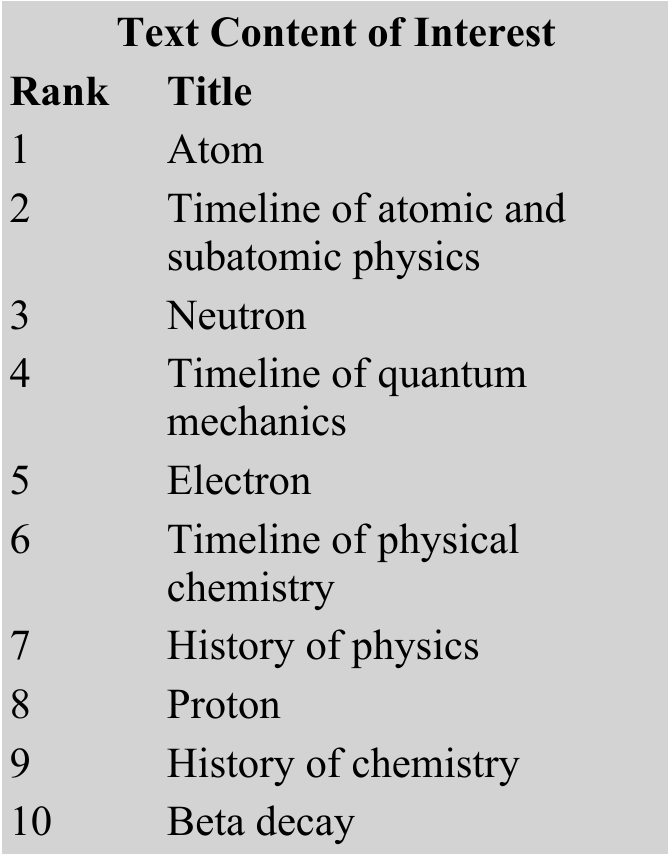}
  }
  \caption{Extract from one experiment to illustrate a reading task
    with subsequent document retrieval:
    (a)~Our data acquisition setup with one participant wearing an EEG
    cap with embedded electrodes. (b)~Sample text with 
    the first two sentences from the Wikipedia document ``Atom''
    (relevant document) and the document 
    ``Money'' (irrelevant document). The color of the words shows the
    explicit relevance judgments by the user (red: relevant; blue:
    irrelevant). The crossed-out words were lost because of
    too much noise in the EEG (e.g., because of eye blinks). The
    framed words ``matter'' and ``atomic'' were the top words predicted to
    be relevant by the EEG-based classifier. Colors and markings were
    not shown to the user. (c)~The top-10 retrieved
    documents, based on the predicted relevant words, are highly
    related to the relevant topic ``Atom''.\label{fig:mindir-demo}}
\end{figure*}

\section{Brain-relevance paradigm for information filtering} %

We propose a new paradigm for information filtering based on
brain activity associated with relevance. The \textit{brain-relevance
  paradigm} is based on the following four hypotheses evaluated
empirically in this paper:

\medskip

\begin{description}
  \item [H1:] Brain activity associated with relevant words is
    different from brain activity associated with irrelevant words.
  \item [H2:] Words can be inferred to be relevant or irrelevant based
    on the associated brain activity.
  \item [H3:] Words inferred to be relevant are more informative for
    document retrieval than are words inferred to be irrelevant.
  \item [H4:] Relevant documents can be recommended based on the inferred
    relevant and informative words.
\end{description}

The following two sections provide the cognitive neuroscience and the
information science motivations as well as existing foundations of the
brain-relevance paradigm.

\paragraph{Cognitive neuroscience motivation.} %
Event-related potentials (ERPs) are obtained by synchronizing
electrical potentials from EEG to the onset
(``time-locked'') of sensory or motoric events. The last 50 years of
psychophysiology have demonstrated beyond a reasonable doubt that ERPs
have a neural origin, that mental events can reliably elicit them, and
that the measurement of their timing, scalp distribution
(``topography''), and amplitude can be invaluable in providing
information on normal~\cite{Kok:1997} and neuropathological
functioning~\cite{Pfefferbaum+:1984}.

Mentally controlling interfaces through measured ERPs has, to
date, principally relied on the P300. The P300 is a distinct, positive
potential that occurs at least 300~ms after stimulus onset and is
traditionally obtained via so-called oddball paradigms. Sutton et
al.~\cite{Sutton+:1965} presented a fast series of simple stimuli with
infrequently occurring deviants (e.g.~1~in~6tones having a high
pitch) and discovered that these rare ``oddballs'' would on average
trigger a positivity compared to the standard stimuli. Later experiments
showed that the degree to which the stimulus provided new
information~\cite{Sutton+:1967} and was
  task-relevant~\cite{Squires+:1977} amplified the P300, whereas
  repetitive, unattended~\cite{Hillyard+:1983} or easily
processed~\cite{Donchin+:1988} stimuli could remove the P300 entirely.

For the language domain, the onset of words normally evokes a
negativity at ca.~400 ms which has been attributed to semantic
processing~\cite{Kutas+:1984}. This N400 was first observed as a type
of ``semantic oddball'' since the closing word in a sentence such as
``I like my 
coffee with milk and torpedoes'' is semantically improbable, but would
amplify the N400 rather than cause a P300.
However, if a rare syntactic violation 
occurs in a sentence (``I likes my coffee [..]''), the deviant word once
again evokes a positivity, but now at 600 ms~\cite{Hagoort+:1993}. As
this P600 shows similarities to the P300 in polarity and topography,
it started the ongoing debate as to whether it is a language-specific
``syntactic positive shift'', or a delayed P300~\cite{Muente+:1998,Osterhout+:1996,
  Sassenhagen+:2014}.   
Finally, research on memory has identified a late
positive component (LPC) at a latency similar to the P600. The LPC
has been related to semantic priming and is particularly strong in
tasks where an explicit judgement on whether a word is old or new is
to be made~\cite{Rugg+:1998}. Consequently, it is often associated
with mnemonic operations such as
recollection~\cite{Paller+:1992}. In the present context, relevant
words could cue recollection of the user's intent, thereby
amplifying the LPC.

Although the P300/P600 and N400 are often described as contrasting
effects, this is not necessarily the case in predicting term
relevance. That is, if an odd, task-relevant stimulus yields a P300
or P600 and a semantically irrelevant stimulus an N400, it follows
that the total amount of positivity between an estimated 300 and 700
ms may 
indicate the summed total semantic task relevance. This was indeed
found by Kotchoubey and Lang~\cite{Kotchoubey+:2001}, who showed that
semantically relevant 
oddballs (animal names) that were randomly intermixed amongst words from four
other categories evoked a P300-like response for semantic relevance
(but at ca.~600~ms). Likewise, our previous work on inferring
term-relevance from event-related potentials~\cite{Eugster+:2014},
showed that a search category elicited either P300s/P600s in response
to relevant words or N400s evoked by semantically irrelevant terms.

\paragraph{Information science motivation.} %
Relevance estimation aims to quantify how well the retrieved
information meets the information need of the user. Computational
methods are used in estimating statistical relevance measures based on
word occurrences in a document collection. These measures are used in
many information retrieval applications, such as Web search engines,
recommender systems, and digital libraries. One of the most well-known
statistical measures of word informativeness or word importance is
\emph{tf-idf}~\cite{Sparckjones:1972}.

The foundation of \emph{tf-idf} is that low- and medium frequency-words
have a higher discriminating power at the level of the document
collection, in particular when they have high frequency in an
individual document \cite{Sparckjones:1972}. For example, the word
``nucleus'' has a low frequency at the collection level but a higher
frequency in a document about atoms (i.e., the ``Atoms'' document) and
therefore is considered to 
discriminate this document better than, for example, the word ``the,''
which has a high frequency at both the collection and document levels.
Search and recommendation systems use word-importance statistics to
produce a ranked list of documents that match 
the word list encoding the user's search intent. The words in
documents can be indexed with weights encoding their importance, and
ranking models 
then compute a relevance score for each document in the document
collection and rank the documents according to the relevance
scores. For example, if the words ``the'' and ``nucleus'' are encoding
the user's intent, then a ranking model could estimate
that the document ``Atom'' should be ranked higher because it has a high
importance value for the word ``nucleus'' compared to, for example,
the document ``Nuclear Magnetic Resonance,'' which also contains the 
words ``nucleus'' and ``the,'' but with lower importance.

In summary, word relevance is determined by the user given the user's
search intent. Word informativeness is determined by the search
system given the document collection. Words that are both relevant and
informative are words that discriminate relevant documents from
irrelevant documents and are needed to recommend meaningful documents.
In addition to the brain-activity findings related to the semantic
oddball (introduced in the cognitive neuroscience motivation), recent
findings in quantifying brain activity associated with language 
also suggest a connection between the word class and frequency of the
word, and the corresponding brain activity. It has been shown that
brain activity is different for different word classes in
language~\cite{Munte200191} and that high-frequency words elicit
different activity than low-frequency words~\cite{Hauk20041090}.

\section{Methodology}

During the experiment, we recorded the EEG signals of 15 participants
while each participant performed a set of eight reading
tasks. Experimental details are provided in
SI~Neural-Activity Recording Experiment.

\begin{figure*}[t]
  \centerline{
\raisebox{4cm}{\textbf{a}}
\includegraphics[height=1.5in]{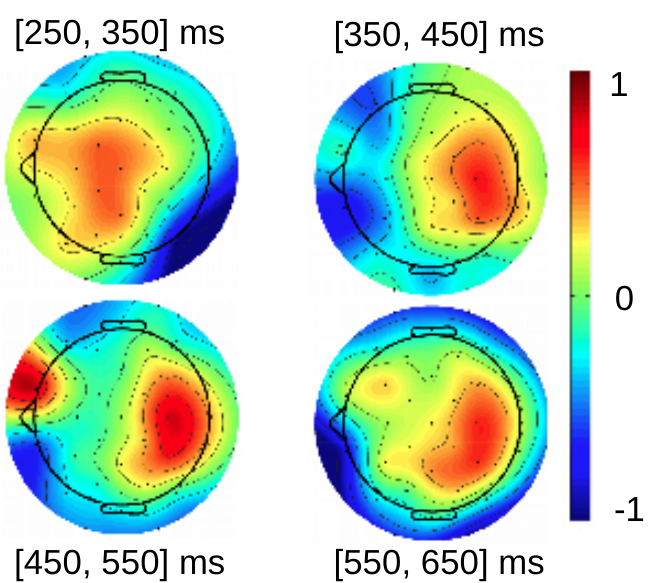} \quad
\raisebox{4cm}{\textbf{b}}
\includegraphics[scale=1]{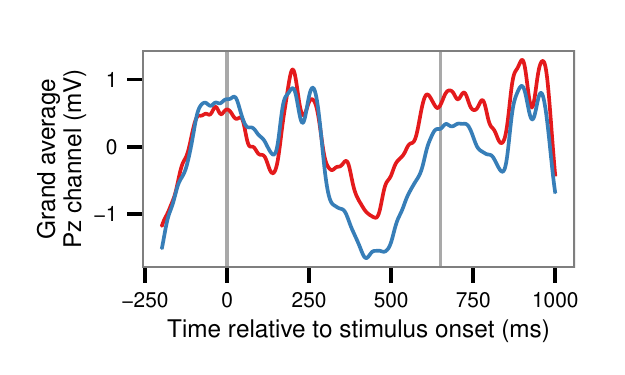}
\raisebox{4cm}{\textbf{c}}
\includegraphics[scale=1]{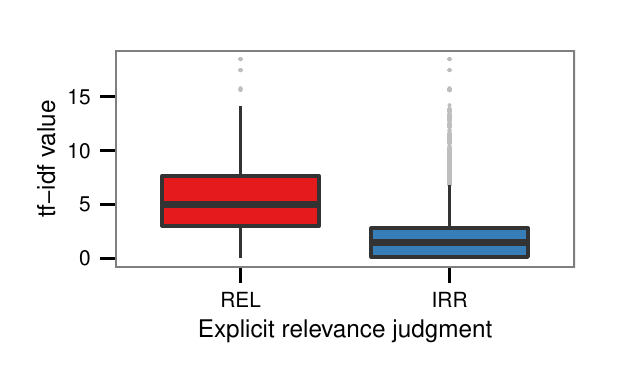}
  }
  \caption{Grand average results over all participants and reading
    tasks based on explicit relevance judgments: %
    (a)~Grand average-based topographic scalp plots of relevant minus
    irrelevant ERPs from $[250, 350]$~ms, $[350, 450]$~ms, $[450,
    550]$~ms, and $[550, 650]$~ms after word onset. %
    (b)~Grand average event-related potential at the Pz
    channel of relevant (red curve) and irrelevant (blue curve)
    terms. The gray vertical lines show the word onset events.
    (c)~Term frequency--inverse document frequency values
    (\textit{tf-idf}) of relevant 
    (red box plot) and irrelevant (blue box plot) words. The median
    of relevant words is $5.00$, and that of irrelevant words is
    $1.46$. The difference is significant (Wilcoxon test, $V =
    49680192$, $p < 0.0001$).\label{fig:grand-results}}
\end{figure*}

\paragraph{Reading task.} %
The text content read by the user
consisted of two documents at a time. Each document was chosen from
a list of 30 candidate documents, and each document was selected from a
different topical area. For example, the documents ``Atom,''
``Money,'' and ``Michael Jackson'' were part of the list; SI~Table~1
provides a detailed list of the documents. One document represented
the \textit{relevant topic}, the other one an  
\textit{irrelevant topic}. The user chose the relevant topic herself
in the beginning of the experiment. The user read the first six
sentences from each document---first the first sentence from both
documents, then the second sentence from both documents, and so
on. The obtained term-relevance feedback (predicted from brain
signals) was then used to retrieve further documents relevant to the
user-chosen topic of interest, from among the four~million documents
available in the database.

In order for the task to be representative of natural reading, no
simplifications were done on the text content. In particular, this
implies that the sentences have different numbers of words, and word
length ranges from very short to very long. Figure~\ref{fig:mindir-demo}
illustrates one reading task consisting of the relevant document
``Atom'' and the irrelevant document ``Money'' with subsequent
document retrieval.

\paragraph{Data analysis.}
To associate brain activity with relevance, we computed
the neural correlates of relevant and irrelevant words for all
participants.  A participant-specific 
single-trial prediction model~\cite{Blankertz+:2002} was computed
for each participant, and the performance was evaluated on a left-out reading
task (leave-one-task-out, a cross-validation scheme). This procedure
matches the example of the task illustrated in
Figure~\ref{fig:mindir-demo}, consisting  
of the following steps: (1)~users perform a new reading task;
(2)~relevance predictions are made for each word based on a model that
was trained on observations collected during previous reading tasks;
and (3)~documents are retrieved using the relevance
predictions for the present reading task.  We present results for the
(H1)~neural correlates of relevance, (H2)~term-relevance prediction, 
(H3)~relation between relevance prediction and word importance, and
(H4)~document recommendation. Results are presented for both individual
users and as grand averages. Technical details are in SI~Data
Analysis Details.

\paragraph{Evaluation.}
To quantify the significance and the effect sizes of the
\textit{brain feedback}-based prediction performances, we compared
them against performances from prediction models learned from
\textit{randomized feedback}.  By comparing 
against this baseline, we are able to operate with natural and hence
non-balanced texts. Standard permutation tests~\cite{Good:2000} were
applied for significance testing.

We used the \textit{Area under the ROC curve} (AUC) to quantify the
performance of the classifiers. AUC is a widely used and sensible
measure even under the class imbalances of our scenario, and it is a
comprehensive measure for comparison against the prediction models
based on randomized feedback.
From the perspective of document recommendations, it is more important
to predict relevant words than to predict irrelevant words. To
quantify this, we measured the \textit{precision}~(SI~Appendix,
Equation~1). To demonstrate the influence of a positive
predicted word on the document retrieval problem, we additionally
measured the \textit{tf-idf-weighted precision}~(SI~Appendix,
Equation~2).
From the user perspective, the quality of the recommended documents is
important. To quantify this, we used \textit{cumulative imformation
  gain}, which measures the sum of the graded relevance values of the
returned documents~(SI~Appendix, Equation~7).
AUC and precision are based on participant-specific relevance
judgments, and cumulative information gain is based on external
topic-level expert judgments. Details on the concrete definition of
the evaluation measurements and the assessment process are available
in the SI~Appendix.


\section{Results}

\paragraph{Neural correlates of relevance.}
Grand-average based ERP results show that brain activity
associated with relevant words is different from brain activity
associated with irrelevant words~(H1), over all participants and
all reading tasks. The topographic scalp plots in
Figure~\ref{fig:grand-results}a show the spatial interpolation of
relevant ERPs minus irrelevant ERPs over all electrodes from 300ms to
600ms after a word was shown on screen.
The topography of the difference showed an initial fronto-central
positivity at 300ms, relative to the onset of the word on the screen,
followed by a centro-parietal positivity from 400 to 600 ms. The
maximal effect of relevance can be clearly observed in
Figure~\ref{fig:grand-results}b, with $-0.24 \mu$V for relevant words
and $-1.06 \mu$V for irrelevant words at 367ms over Pz. Following the
negativity a late positivity can be observed for both types of words,
which reaches a local maximum at a latency of around 600 ms,
implicating a possible P600 or LPC.

For descriptive purposes, we tested the difference between the relevant
and irrelevant words of well-known P300, N400, and P600 ERP components 
and their latencies given in the existing literature.
There was no significant difference in the early P3 interval ($[250,
350]$~ms, paired $t$-test, $T(14) = 1.75$, $p =
0.10$), which suggests that the system does not rely
on the mere visual resemblance between relevant words and the intent
category.
However, irrelevant words elicited a negativity compared to
relevant words in the N400 window ($[350, 500]$~ms, $T(14) = 2.27$, $p
= 0.04$). Moreover, relevant words  were found to significantly elicit
a positivity compared to irrelevant words in the P600 interval ($[500,
850]$~ms, paired $t$-test interval, $T(14) = 4.99$, $p = 0.0002$).
For the purpose of the subsequent term-relevance prediction, this
result verified our approach of computing the temporal features for the
ERP classification within the range of 200ms to 950ms (this range was
determined based on the pilot experiments).
SI~Figure~1 shows the remaining scalp plots for other time intervals,
and SI~Figure~2 shows the grand-average-based ERP curves for all
channels.

\begin{figure}[t]
  \centerline{
\includegraphics[scale=1]{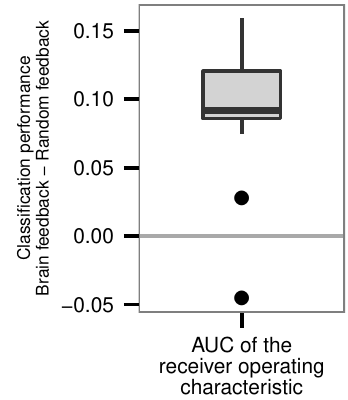}
\includegraphics[scale=1]{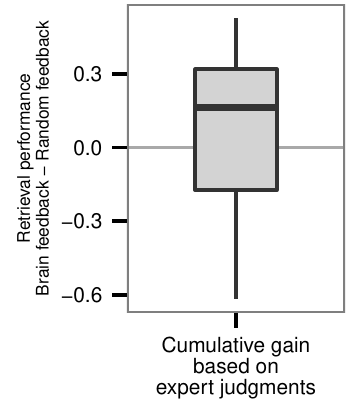}
  }
  \caption{
    Overall prediction and retrieval performances:
    (a)~Overall classification performance on new data, measured by the difference
    of AUC between a classifier learned using explicit relevance
    feedback and that learned using randomized feedback. The difference is
    significantly greater than zero ($p < 0.0005$). The figure shows
    that the prediction models are able to find structure
    significantly discriminating 
    between relevant and irrelevant brain signal patterns. %
    (b)~Overall retrieval performance characterized as difference in
    cumulative gain (based on expert judgments) between
    documents retrieved based on brain-based feedback and randomized
    feedback (normalized with the maximum information gain that would
    be possible to achieve when retrieving the best top-30
    documents). Brain feedback is significantly better than randomized 
    feedback ($p < 0.003$).\label{fig:classif-perf}}
\end{figure}

\begin{figure}[t]
  \centerline{
    \includegraphics[scale=0.94]{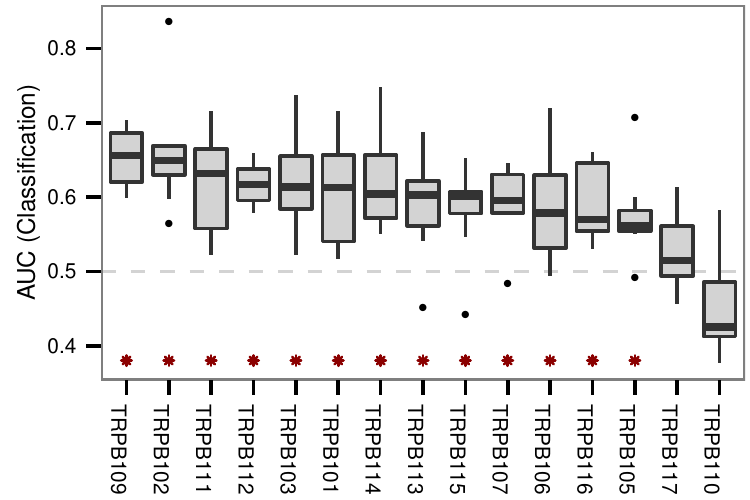}
  }
  \caption{Comprehensive term-relevance prediction performance on participant
    level: classification performance for all participants (TRPB\#), measured
    as AUC for participant-specific models on left-out reading
    tasks. The horizontal dashed line indicates the performance of a
    model learned using randomized feedback. Asterisks indicate models
    with significantly better AUC ($p < 0.05$; exact
    $p$-values are listed in SI~Table~3).\label{fig:classif-auc}}
\end{figure}

\begin{figure}[t]
  \centerline{
    \includegraphics[scale=0.94]{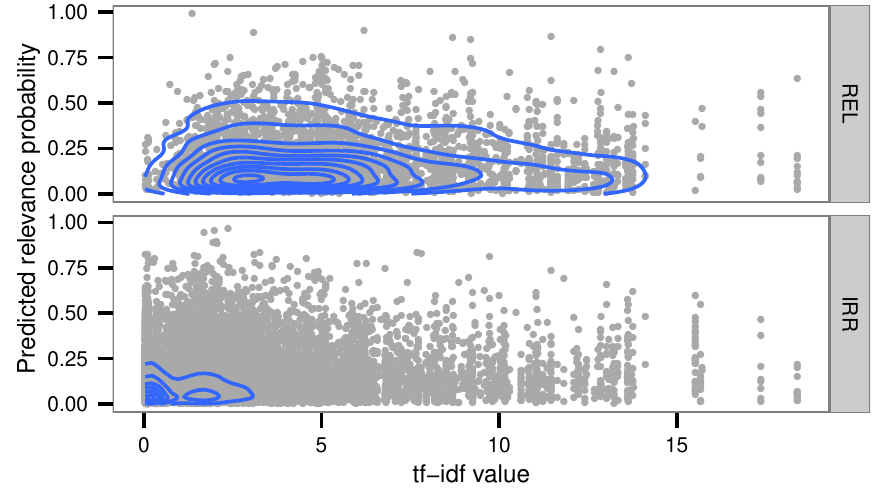}
  }
  \caption{Relevance prediction versus tf-idf value: Two-dimensional
    kernel density estimate (the blue contours) for relevant (top) and
    irrelevant (bottom) words with an axis-aligned bivariate normal
    kernel. The mass of relevant words (REL) is much more towards the top
    right corner (high probability to be relevant and high tf-idf
    value) than the mass of irrelevant words (IRR). The gray points in the
    background are the observed words.\label{fig:classif-tfidf}}
\end{figure}

\begin{figure}[t]
  \centerline{
    \includegraphics[scale=0.94]{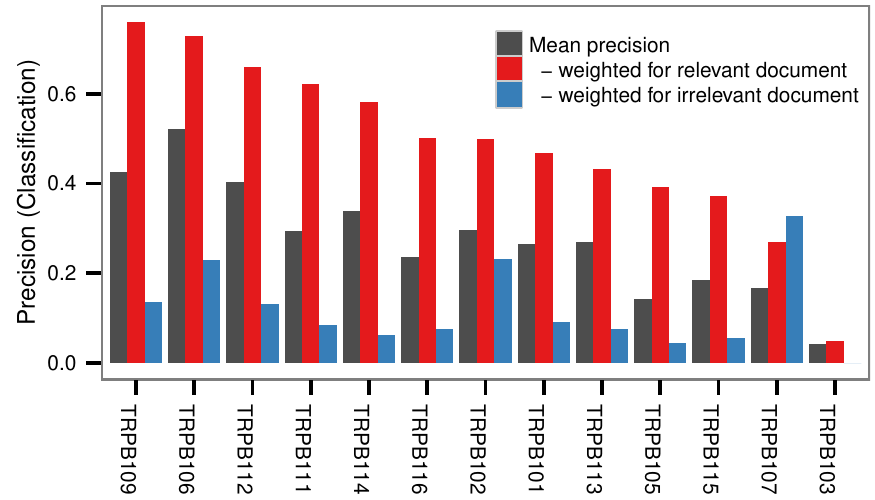}
  }
  \caption{Relevance prediction weighted for retrieval on participant
    level: Mean precision per participant (black bars). For the
    document retrieval problem, the influence of a positive
    predicted word is dependent on its document-specific \textit{tf-idf}
    value. Therefore, a false positive can have a smaller effect than
    a true positive. The red and blue bars illustrate this
    effect. The red bars show the precision weighted with the
    \textit{tf-idf} value of the relevant document. The blue bars
    show the precision weighted with the \textit{tf-idf} value of the 
    irrelevant topic.\label{fig:classif-prec}}
\end{figure}

\begin{figure}[t]
  \centerline{
    \includegraphics[scale=0.94]{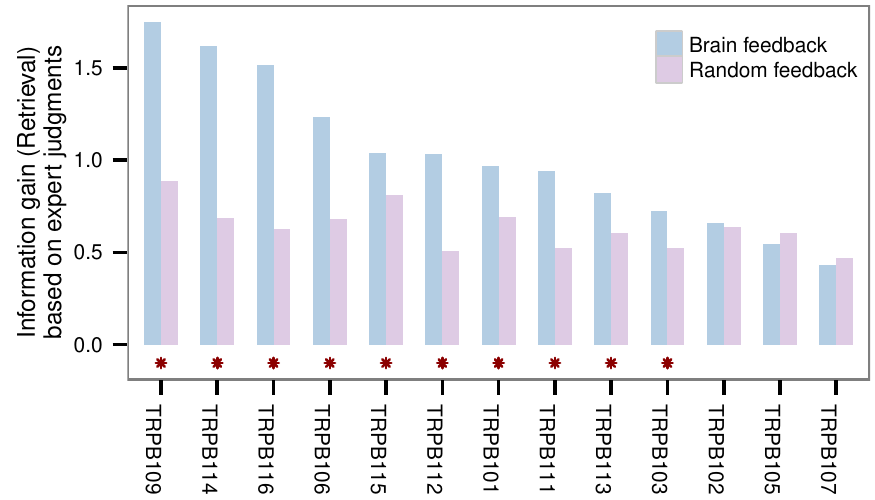} %
  } %
  \caption{Retrieval performance of each participant: %
    Average cumulative information gain (on a scale of 0-3) based on the
    top-30 retrieved documents for the participant. Asterisks indicate
    a significantly better pooled information gain based on brain
    feedback than random feedback retrieval based on
    1000~iterations ($p < 0.05$; exact $p$-values are listed in
    SI~Table~4).\label{fig:retriev-cumgain}} 
\end{figure}

\paragraph{Term-relevance prediction.}
Across participants and reading tasks, the classification of brain
signals by models learned from earlier explicit feedback shows
significantly better results than with models learned from randomized
feedback (Figure~\ref{fig:classif-perf}a; $p < 0.0005$,
Wilcoxon test, $V = 118$). This implies that the prediction models
are able to extract and utilize structured signals significantly and
that words can be inferred to be relevant or irrelevant based on the 
associated brain activity~(H2).

Figure~\ref{fig:classif-auc} shows the classification performance in
terms of AUC for each participant. For 13 out of the 15
participants, the term-relevance prediction models perform
significantly better than does a prediction model learned based on
randomized 
feedback (hence having $\text{AUC} = 0.5$; $p < 0.05$,
within-participant permutation tests with $1000$ iterations). For two
participants, the predictions were essentially random, and they were
excluded from the rest of the analyses. It is well known that BCI
control does not work for a non-negligible portion of participants
(ca.~15-30\% \cite{Vidaurre:2010}), and the reported results should
be interpreted as being valid for the population of users, which can be
rapidly screened by using the system on pre-defined tasks.

\paragraph{Relevance for document retrieval.}
For our final goal, the retrieval and recommendation of documents, it
is important to be able to detect words that are both relevant and
informative (measured by the \textit{tf-idf}) in discriminating between
relevant and irrelevant documents in the full collection.
Figure~\ref{fig:classif-tfidf} visualizes the relationship between
the predicted relevance probability of words and their \textit{tf-idf}
values. Relevant words (according to the user's own judgement
afterwards) are predicted as being more relevant than irrelevant
words, but also, their \textit{tf-idf} values are greater~(H3). 
The figure further indicates that the \textit{tf-idf} dimension
explains more of the difference than does the predicted relevance.

In terms of an information retrieval application, the precision of the
prediction models is the most important measure. For document
retrieval, the influences of positive predicted words on the search
results are not equal but rather dependent on the word-specific
\textit{tf-idf} values within each 
individual document. For example, a true positive predicted word
can still have very low impact on the search result if its
tf-idf value is low in the relevant
documents. Similarly, a false positive predicted word can have
only a low impact on the search result, if its \textit{tf-idf} value
is low. Figure~\ref{fig:classif-prec} visualizes the mean precision of
the prediction models from the perspective of the retrieval problem. It
shows the mean precision for each of the 13 participants over all
of the reading tasks (based on binarizing the predicted probabilities with
the threshold $0.5$). In addition, it shows what is actually
crucial: The precision weighted with the \textit{tf-idf} values from
the relevant document is in all cases, except for one, much higher
than the precision weighted with the \textit{tf-idf} values from the
irrelevant document.

In conclusion, the results in Figure~\ref{fig:classif-prec} explain
why the prediction models are useful for document retrieval and
recommendation, even though the unweighted precision of the prediction
models is limited. In detail, our prediction models tend to predict
true positive words with higher \textit{tf-idf} values, and false
positive words with lower \textit{tf-idf} values. This means that our
prediction models tend to predict words that the
user would judge to be relevant, and which are also
discriminative in terms of the user's search intent.

\paragraph{Document recommendation.}
The final step is to use the relevant words---predicted from brain
signals---for document retrieval and recommendation, and to evaluate
the cumulative gain. Figure~\ref{fig:classif-perf}b
shows that across the participants and reading tasks, document
retrieval performance based on brain feedback is significantly better
than randomized feedback (top-30 documents, $p < 0.003$,
Wilcoxon test, $V = 3153$). Therefore, relevant documents can be
recommended based on the inferred relevant and informative words~(H4).

Figure~\ref{fig:retriev-cumgain} shows the document retrieval
performance for each participant in terms of mean information
gain. Based on the expert scoring, the scale for the mean information
gain is from 0 (irrelevant) to 3 (highly relevant). The visualization
shows for each participant the mean information gain over all reading
tasks based on brain feedback (blue bars) and randomized feedback
(purple bars). For 10 participants, the brain feedback results in
significantly greater information gain ($p < 0.05$;
two-sided Wilcoxon test). SI~Figure~3 also shows the visualizations
for top~10 and top~20 retrieved documents. In both cases, the same
significant results hold except for one participant~(TRPB113).

\section{Discussion}

By combining insights on information science and cognitive
neuroscience, we proposed the brain-relevance paradigm to construct
maximally natural interfaces for information filtering: The user just
reads, brain activity is monitored, and new information is
recommended. To our knowledge, this is the first end-to-end
demonstration that the recommendation of new information is possible
just by monitoring the brain activity while the user is reading.

The brain-relevance paradigm for information filtering 
is based on four hypotheses empirically demonstrated in this paper. We
showed that (H1)~there is a difference in brain activity associated
with relevant versus irrelevant words; (H2)~there is a 
difference in the importance of words depending on their relevance
to the user's search intent; (H3)~it is possible to detect
relevant and informative words based on brain activity; and (H4)~it is
possible to recommend relevant documents based on the detected
relevant and informative words.

From a cognitive neuroscience point of view, it is known that specific
ERPs can be particularly associated with relevance. In cognitive
science, early P300s have been related with task
relevance. 
In psycholinguistics, N400s are commonly associated
with semantic processes~\cite{Kutas+:1984} as semantically
incongruent words amplify the component whereas semantic relevance
reduces it. Late positivity has been related to semantic
task-relevant stimuli~\cite{Kotchoubey+:2001}, in particular if
characterizing it as a 
delayed P3 response, due to the assessing of relevance of language,
or an LPC, due to mnemonic operations and semantic judgements.
In line with these findings, our grand averages indicate that the ERP
at a latency of 500--850 ms is most likely the best predictor of
perceiving words that are semantically related to a user's search
intent. 
The present data do not allow for a dissociation among
the P300, N400 or P600 as the most likely neural candidate for evoking
the observed effect. Indeed, the method is based on the assumption that
task relevance and semantic relevance both contribute positively to the
inference of relevance when aiming to ultimately predict a user's search
intent without requiring an additional task by the user.

While our results use real data and are also valid beyond the
particular experimental setup, our methodology is limited to
experimental setups in which it is possible to control strong noise,
such as noise due to physical movements, which are known to cause
confounding artifacts in the EEG signal. Another limitation is that
the comparison setup in our studies considers only two topics at a
time, one being relevant and another being irrelevant. While this is a
solid experimental design and can rule out many confounding factors,
it may not be valid in more realistic scenarios in which users choose
among a variety of topics during their information seeking activities.
Furthermore, the presented term-relevance prediction
is based on a traditional set of event related potentials to
demonstrate the feasibility of the methodology. However, it is
possible that more advanced feature extraction could improve the
solution further, for example by computing phase synchronization
statistics in the delta and theta frequencies, which recently have
been shown to be sensitive to the detection of relevant lexical
information~\cite{Brunetti+:2013}.

Despite these limitations, our work is the first to address an
end-to-end methodology for performing fully automatic information
filtering by using only the associated brain activity. Our
experiments demonstrate that our method works without any requirements
of a background task or artificially evoked event-related potentials;
the users are just reading text, and new information is
recommended. Our findings can enable systems that analyze relevance
directly from individuals' brain signals naturally elicited
as a part of our everyday information seeking activities.

\section{Materials}

The SI~Appendix provides extensive details on all technical aspects.
SI~Database describes the selection process and the criteria for the
pool of candidate documents.
SI~Neural-Activity Recording Experiment provides the experimental
details, i.e., the participant recruiting, the procedure and design of
the EEG recording experiment, the apparatus and stimuli definition,
and details on the pilot experiments.
SI~Data Analysis Details describes the general prediction evaluation
setup, EEG cleaning and preparing, and the EEG feature engineering.
SI~Term Relevance Prediction gives a description of the Linear
Discriminant Analysis (LDA) method used for the prediction models, and
specifics on the evaluation measures for prediction.
SI~Intent Modeling based Recommendation gives details on the intent
estimation model based on the LinRel algorithm, and specifics on the
evaluation measures for document retrieval.

\section{Acknowledgments}

We thank Khalil Klouche for designing Figure~1.
This work has been partly supported by the Academy of Finland (278090;
Multivire, 255725; 268999; 292334; 294238; and the Finnish Centre of
Excellence in Computational Inference Research COIN), 
and MindSee (FP7 ICT; Grant Agreement 611570).


\clearpage

\twocolumn[{
\begin{center}
{\LARGE Supplemental Information SI\\[0.5em]
Natural brain-information interfaces: Recommending information by
relevance inferred from human brain signals\\[2em]}
\end{center}
}]

\section{SI Database}

The database used in the experiment was the English Wikipedia provided
by Wikimedia (database dump of 2014/07/07 \cite{WPDump}). For the
experiment, our search engine indexed all articles except special
pages such as disambiguation pages. The references, notes, and external
links were removed from the text of the articles. The final
database contained over 4~million articles.

The \textit{pool of candidate documents} read by the participants
during the experiment consisted of 30 documents. The criteria for
choosing a document were that (1)~the document should describe
a topic of general interest and that (2)~the first six sentences of the
introduction of the document provide a sufficient description of the
topic. The final pool of documents fulfilling these criteria are listed
in SI~Table\ref{tab:document-stats}.

\section{SI Relevance Judgments of Words and Documents}

In order to measure the relevance prediction performance, ``ground
truth'' in the form of relevance judgements for individual words is
needed, for a specific reading task, on both relevant and irrelevant
document. The binary relevance judgment ``relevant'' or
``irrelevant'' of a word was provided by each participant during the
experiment (see SI Neural Activity Recording Experiment). This allowed
us to capture the subjective nature of perceived relevance. In
addition, for each document, the word class of each word was defined
(nouns, verbs, adjectives, etc.) and three experts judged each word as
being ``relevant'' or ``irrelevant'' for the given document.

In order to measure the document retrieval performance, the
``ground truth'' relevance judgments of retrieved documents given a
relevant topic of the reading task is needed. For each of the 30
documents in the pool, three experts judged all documents that were
retrieved in any experiments (brain feedback-based and random
feedback-based), 
resulting in a pool of 13971 retrieved documents. The experts
assessed all the documents according to the following criterion:
``Would you be satisfied in having this document in the search result
list of documents after examining document \textit{x}? If yes, how
satisfied from 1 to 3, if no 0.'' The mean Cohen's Kappa
\cite{Cohen:1968} indicated substantial agreement between the experts,
$\text{Kappa} = 0.72$.

\section{SI Neural Activity Recording Experiment}

We recorded the electroencephalography (EEG) signals of 17
participants while each participant performed a set of eight reading
tasks. The following sections provide the experimental details.

\subsection{Participants}

Participants were volunteers recruited from the universities of the
Helsinki metropolitan area in Finland. They were selected only if they
were right-handed, had no self-reported neuropathological history, and
were deemed to have sufficient fluency in English. Handedness was
assessed using the Edinburgh Handedness Inventory~\cite{Oldfield:1971,
  RightHandTest} and English fluency using the Cambridge English ``Test your
English -- Adult Learners'' online test~\cite{EnglishTest}. Seventeen
participants were recruited to 
participate in the experiment. The data of two participants were
discarded due to 
technical issues. Of the fifteen remaining, 8 were female and 7
male. Their English fluency was assessed as high ($\text{Mean} =
23.53$, $\text{SD} = 1.23$), 
and their handedness as right-handed ($\text{Mean} = 87.35$, $\text{SD}
= 12.13$). They were 
fully briefed as to the nature and purpose of the study prior to the
experiment. Furthermore, and in accordance with the Declaration of
Helsinki, they signed informed consents and were instructed on their
rights as participants, including the right to withdraw from the
experiment at any time without fear of negative consequences. They
received two movie tickets as compensation for their participation.

\subsection{Procedure and Design}

Following the initial briefing, participants were explained the task in
more detail, while the EEG equipment was set up. They then received a
short training task with two sample topics. When participants indicate their
complete understanding of the task, the experiment
commenced. Participants completed eight experimental blocks, each
consisting of a single reading task with two topics, drawn randomly
(without replacement) from the pool of 30 document candidates. At the
beginning of the block, they were asked to freely choose which of the
two documents described the relevant and irrelevant topic. Every block
comprised six trials, each consisting of one sentence from the
relevant and one sentence from the irrelevant document, with the
presentation order of the sentences randomized between blocks. Each
trial consisted of the sequential presentation of words (the word
stream), two validity sub-tasks, and an explicit word relevance
judgment task (judging the words as ``relevant'' or ``irrelevant'').

Every trial started with a warning signal (the words ``Starting
trial''), followed by the presentation of the mask. An initial sentence
separator was shown before the word stream was shown. The word stream
consisted of the sequential presentation of each word in the first
sentence, followed by a sentence separator, the words in the second
sentence, and concluded by a final sentence separator. Every word and
sentence separator was presented for exact 699~ms ($\text{SD} =
0.3\text{ ms}$). 
Punctuation marks were not shown.
Masking effects were countered to some extent by the
frame resizing, which keeps the level of foveal stimulation
constant. Prior tests suggested that people had more difficulty
reading with than without short masks between the bursts, so as a
consequence we removed them. It is possible these masking effects
may be much more significant with strong "flashing", as would be the
case with very short stimulus durations. Here, the words appearing
at a slow rate of ca 700 ms per words made reading very easy.

Following the word stream, two extra sub-tasks were presented to
validate that the participants had remembered their chosen word and that
they had paid attention to both sentences. First, they were asked to
type in the name of 
the relevant topic in order to ascertain they had not forgotten. Then,
a recall task was presented to prevent the participants from selectively
concentrating on one of the two sentences. One of the sentences was
selected randomly and presented in full on the screen, with one of the
nouns or verbs substituted by question marks. Participants were asked
to type in the word missing in the sentence. They were then presented
with feedback in points regarding their performance on these two tasks
as a motivational instrument (similar to \cite{Spape+:2011}).

Then, in the final part of the trial, the participants were asked to
explicitly rate the relevance of all words from the relevant
topic. All words 
were shown in one (if the sentence comprised fewer than 35 words) or
two columns on the screen. A cursor was presented next to each word,
indicating a two-alternative forced-choice decision. Pressing the left
arrow key on the keyboard would rate the word as irrelevant and
pressing the right would rate it as relevant. Participants were
instructed prior to 
the experiment that they should not re-interpret the relevance of the
words and instead make a decision based on their previous viewing of
the sentence. To facilitate this, they received a maximum of 2~s to
respond to each word, after which the cursor moved to the next word in
the sentence. After the last word was rated, the trial was completed,
with the next trial starting after an inter-trial interval of ca. 1~s,
unless it was the last trial in the block.

After completing a block, they were requested to freely write about
their chosen, relevant topic; this task was defined to keep the
participant engaged. Finally, they filled out a questionnaire
with two items for both topics, one regarding their interest (``how
interesting do you find topic \textit{x}'') and one regarding their
knowledge (``how much do you know about topic \textit{x}'') using a
9-point rating scale (1: not at all -- 9: extremely so). Three
self-timed breaks with a minimum of one minute evenly split the
blocks into four parts. The experiment, excluding preparation and
instruction, lasted approximately one hour.

\subsection{Apparatus and Stimuli}

Words were presented with an 18-point Lucida Console black typeface at
the center of the 19'' LCD screen. They were shown against a silver
(RGB $82\%$, $82\%$, $82\%$) background in the middle of a $300
\times 100$ pixel pattern mask. The mask was a black
rectangle with a grid-like pattern, with an opening to show the
word. This was used to control the degree to which word length
affected light reaching the eyes (i.e. to make sure longer
words were not tantamount to more black pixels on the
screen). Sentence separators were word-like character
repetitions consisting of 4 to 9 numbers (\verb|3333333|) or other
non-alphabetic characters (\verb|&&&&&&|), which were designed to
mimic the same 
early visual activity as words without evoking
psycholinguistic processing.

The screen was positioned approximately 60 cm from the
participants and was running at a resolution of 1680 x 1050 and a
refresh rate of 60 Hz. Stimulus presentation, timing, and EEG
synchronization were controlled using E-Prime 2 Professional 2.0.10.353
on a PC running Windows XP SP3. EEG was recorded from 32 Ag/AgCl
electrodes, positioned on standardized (using EasyCap elastic caps,
EasyCap GmbH, Herrsching, Germany), equidistant electrode sites of the
$10-20$ system via a QuickAmp (BrainProducts GmbH, Gilching, Germany)
amplifier running at 200~Hz. Additionally, the electro-oculogram for
vertical eye movements (and eye blinks) and horizontal eye movements
was recorded using bipolar electrodes positioned respectively 2~cm
superior/inferior to the right pupil and 1 cm lateral to the outer
canthi of both eyes.

\subsection{Pilot experiments}

Prelimary versions of the final experimental procedure and design were
piloted with four separate participants. In these experiments, we
tested and evaluated, for example, the stimulus duration, the explicit
feedback task, and the points system. The data of these pilot
experiments were not used in the final analysis, except that some basic
parameter estimations for the final feature engineering process were
based on cross-validation experiments on these data (e.g., number of
feature windows).

\section{SI Data Analysis Details}

The evaluation setup for prediction and retrieval followed the general
block structure defined by the experimental design. We applied an
participant-specific and leave-one-block-out learning and evaluation
strategy. The individual prediction models are single-trial prediction
models~\cite{Blankertz+:2002}. We report averaged prediction and
retrieval performance, unless otherwise noted.

In detail, for a given participant, $B = \{1, ..., 8\}$ blocks with
explicit term relevance judgments provided by the participant were
available. In order to retrieve a \textit{brain-feedback}-based list
of relevant documents for a specific block $b$, two steps were
executed. First, to obtain a term relevance prediction model for the
given block $b$, a classification model $f_b$ was trained using the
data from the remaining $\{B \setminus b\}$ blocks. The prediction
performance of $f_b$ was then evaluated on the left-out block
$b$. Second, to retrieve the set of documents for block $b$, the set
of terms predicted to be relevant by the classifier $f_b$ with a
probability higher than $0.5$ were used. The retrieval performance was
evaluated against the expert judgements of document relevance for the
relevant topic of block $b$.

As a baseline comparison, we evaluated the brain feedback-based
performances against \textit{random-feedback}-based performances. The
random-feedback scenario corresponds to standard permutation tests and 
results in permutation-based $p$-values \cite{Good:2000}. The
following sections give concrete details on the methodology used.

\subsection{EEG cleaning and preparing}

The  EEG signals were cleaned and prepared following standard BCI
guidelines \cite{Blankertz+:2011}. 
During recording a hardware low-pass filter at 1000~Hz
was applied.
The continous EEG
recordings were filtered with a 35~Hz FIR1 low-pass filter and a
$0.5$~Hz high-pass filter. The signal was then divided into epochs
ranging from $-250$~ms to $1000$~ms relative to the onset of each
stimulus. Baseline correction was performed on each epoch using the
pre-stimulus period. A simple heuristic was applied to reject invalid
channels and epochs: First, invalid epochs were estimated based on the
epochs' variances ($< 0.5~\mu\text{V}$) and the max-min criterium
($40~\mu\text{V}$). A channel was removed if the number of invalid
epochs was higher than $10\%$ of all available epochs. After removing
all invalid channels, invalid epochs were estimated again and
removed. This data cleaning approach was carried out in order to
eliminate noise and potential confounds by common artifacts such as
eye movements and blinks, as well as artefacts caused by loose
electrodes or a cap that did not fit
perfectly. Table~\ref{tab:eeg-stats} shows the statistics for the
cleaning process for each participant.

\subsection{Feature engineering}

Event-related potentials are characterized by their temporal evolution
and the corresponding spatial potential distributions. We followed 
standard feature engineering procedures to create spatio-temporal ERP
features for classification~\cite{Blankertz+:2011}. For each epoch,
the raw EEG data (after basic cleaning) were available as the
spatio-temporal matrix $X^{m \times t^\prime}$, with $m$ 
channels and $t^\prime$ sampled time points. For each
epoch, the time was divided into $t = 7$ equidistant windows
between 250~ms and 950~ms after the stimulus onset. The number of
windows was chosen based on data recorded during the pilot
experiments. For each channel, the potential values 
within one window were averaged, resulting in the spatio-temporal
matrix $X^{m \times t}$. The final feature representation of one
epoch was the concatenation of all columns into one vector $X^{m
\cdot t}$. And, for a specific block $b$ with $n$ epochs, the full
spatio-temporal feature matrix used for classification was
$\underline{X}^{n \times m \cdot t}$. Note that the number of channels
$m$ and the number of epochs $n$ were participant-specific, as they
were dependent on the EEG cleaning and preparing
procedure. Table~\ref{tab:eeg-stats} shows the concrete numbers for
each participant.

\section{SI Term Relevance Prediction}

We developed term relevance prediction models within the framework of
the linear EEG model~\cite{Parra+:2005} and single-trial ERP
classification~\cite{Blankertz+:2011}. In detail, we utilized Linear
Discriminant Analysis~(LDA, see~\cite{Hastie+:2009}) and
learned linear binary classifiers, which we used to predict class
memberhsip probabilities. The assumptions of the method are that the
observations $X$ have been drawn from two multivariate Normal
distributions $N(\mu_k, \Sigma)$, 
one for the class of ``relevant'' observations, and the other for the
class of ``irrelevant'' observations. For the estimation of
the models 
we used shrinkage LDA, a covariance-regularized LDA with a shrinkage
parameter selected by the analytical solution developed by Sch{\"a}fer
and Strimmer~\cite{Schaefer+:2005}. The choice of this simple method was
based on the many existing successful applications using this method
in the BCI community~\cite{Blankertz+:2011}. In addition, one major
reason is robustness against class imbalance~\cite{Xue+:2008}, an
obvious situation in the proposed paradigm (see also
Table~\ref{tab:eeg-stats} for the relevance class distribution per
participant).

\subsection{Leave-one-block-out evaluation}
For each participant, we
trained a set of eight classifiers. The 
classifer $f_b$ for block $b$ was trained with the epochs from the
other blocks, i.e., with the spatio-temporal feature matrix
$\underline{X}^{n_l \times m \cdot t}_{\{B \setminus b\}}$. The
classifier $f_b$ was evaluated on the epochs from block $b$, i.e., on
the matrix $\underline{X}^{n_t \times m \cdot t}_{b}$. The performance
measures of interest were the \textit{Area under the ROC curve} (AUC),
\textit{precision}, and \textit{tf-idf}-weighted precision.
The AUC is defined as the area under the ROC curve, which links the
true positive rate to the false positive rate. A perfect model has an
AUC of $1$, and a random model has an AUC of $0.5$. AUC is a global
quality measure of the classification model. This measure was
chosen because it allowed us to correctly evaluate the models in the
existing class imablance scenario and because it is a comprehensive
measure for comparison to the random feedback models.
Precision is defined as 
\begin{equation}
\text{tp} / (\text{tp} + \text{fp})\text{,}
\end{equation}
where $\text{tp}$ is the number of true positives (i.e., relevant
words predicted to be relevant) and $\text{fp}$ is the number of
false positives (i.e., irrelevant words predicted to be
relevant). This measure was chosen because we want to have a high
precision (i.e., many correct relevant words) for the document
retrieval step.
Weighted precision is defined as
\begin{equation}
(w_{\text{tp}} * \text{tp}) / (w_{\text{tp}} * \text{tp} +
w_{\text{fp}} * \text{fp})\text{,}
\end{equation}
where $w_{\text{tp}}$ is the sum of the term-frequency--inverse
document frequency (\textit{tf-idf}) values of the true
positive words, and $w_{\text{fp}}$ is the sum of \text{tf-idf} values
of the false positive predicted words. In our case, the
\textit{tf-idf} values either come from the relevant document or the
irrelevant document. For a positive predicted word that is not
available in a document, the \textit{tf-idf} is set to $0$. This
reflects that this word has no influence on the document retrieval.

\subsection{Random feedback evaluation}
For a given block $b$, a classifier was trained and evaluated on data
with permuted relevance judgments. If executed for a large number of
permutations, this random-feedback strategy is a permutation test,
resulting in a permutation-based $p$-value~\cite{Ojala+:2010}. The
null hypothesis of the test assumes that the brain data and the
relevance judgments are independent. A small $p$-value indicates that
the classifier is able to find a significant structure discriminating
``relevant'' and ``irrelevant'' brain signal patterns. For each block,
$k = 1000$ permutations were performed, meaning that the smallest
possible $p$-value is $0.001$~\cite{Good:2000}.

\section{SI Intent Modeling-based Recommendation}

We developed an intent estimation model to predict how relevant each
term the user read is to the topic of interest. This model was
then used to retrieve new documents from the database. The motivation
for the intent model is that the predictions of the term-relevance
model can indicate the relevance to a topical intent, but the
individual words for which the predictions are drawn may not represent
the whole topic. For example, the words ``matter'' and ``neutrons''
are related to the topic ``Atom,'' but would not alone be sufficient
search terms to retrieve information about the topic
``Atom.'' Therefore, these words are used as positive feedback for the
intent model to predict that, for example, the words ``atom,''
``atomic,'' and ``nucleus'' are also relevant for the user given the
positively predicted words ``matter'' and ``neutrons.'' We call the
resulting model the intent model of the user~\cite{Ruotsalo+:2013}.

\subsection{Document representation}

The documents and words are modeled as a term-document matrix $K$ with
$i$ terms and $j$ documents. The term vector $k_i$ indicates the
weight of a stemmed word for each of the documents. The words are
stemmed using the English Porter Stemmer 
\cite{Porter:1997:ASS:275537.275705}, and the stemmed words are
referred to as terms. Before stemming, English stop words were removed
because 
they appeared in the Apache Lucence 4.10 stop word
list\footnote{https://lucene.apache.org/}. We used tf-idf weighting to
account for the frequency and specificity of each term 
\cite{Sparckjones:1972}.

\subsection{Intent model}

The intent model estimates a weight for each term based on the input
from the term-relevance prediction classifier. The feedback from the
term-relevance predictions is denoted as $ r_i \in [0, 1]$ for a subset
of terms indexed by $i$. We assume that the term-relevance prediction
$r_i$ of a term 
$k_i$ is a random variable with expected value $E[r_i]=k_i \cdot w$,
such that the expected weight is a linear function of the terms. The
unknown weight vector $w$ is essentially the representation of the
user's intent and determines the relevance of terms.

To estimate $w$ we utilize the LinRel
algorithm~\cite{Auer:2003:UCB:944919.944941}.  It learns a linear
regression model of the form $r=wK$. LinRel allows control for the
uncertainty related to the term weight estimates. The choice of this
method was based on its robustness against suboptimal input, which is
the case for potentially noisy predictions of the term-relevance
prediction model. 

LinRel computes a regularized regression weight vector for each term
$k_i$ in $K$:
\begin{equation}
 a_i = k_i(K^{\top} K + \lambda I)^{-1} K^{\top},
\end{equation}
where $I$ is the identity matrix, and $\lambda$ is a regularization
parameter set to $0.5$, and all terms except $k_i$ on the
right-hand side are shared for all keywords. Then for each keyword,
the final relevance score ${w}_{i}$ at the current iteration is
computed by taking into account the feedback obtained so far:
\begin{equation}
w_{i} = a_i \cdot s_{t} + \frac{c}{2} \| a_i \|,
\label{formula:explr}
\end{equation}
where $s_{t}$ is the vector of term-relevance predictions obtained,
$a_{i}$ is the weight vector of a single keyword $i$ in the data $K$,
$\| a_i \|$ is the $L_{2}$ norm of the weight vector, and the constant
$c$ is used to adjust the influence of the history (we used $c = 2$ to
give equal weight for exploration and exploitation).  It can be shown
that this procedure is equivalent to estimating the upper confidence
bound in a linear regression problem
\cite{Auer:2003:UCB:944919.944941}.

\subsection{Retrieval model}
 
Intent model estimates a weight $w$ for each term which, in turn, is
used to retrieve new documents from the  database, to be recommended
for the user. We use a unigram language modeling approach of
information retrieval~\cite{Ponte:1998:LMA:290941.291008}.  
In detail,  the vector ${w}$ is treated as a sample of a desired
document, and documents $d_j$ are ranked by the probability that
$w$ would be generated by the respective language model
$M_{d_j}$ for the document $d_j$. 

Using maximum likelihood estimation, we get
\begin{equation}
P(w\vert M_{d_j}) = \prod_{i=1}^{|\mathbf{w}|} \hat{P}_{mle}(k_i\vert M_{d_j})^{w_i},
\end{equation}
and to avoid zero probabilities and improve the estimation we then
compute a smoothed estimate by Bayesian Dirichlet smoothing
so that
\begin{equation}
\hat{P}_{mle}(k_i\vert M_{d_j})
= \frac{c(k_i|d_j)+\mu p(k_i|C)}{\sum_k c(k|d_j)+\mu},
\end{equation}
where $c(k|d_j)$ is the count of term $k$ in document $d_j$,
$p(k_i|C)$ is the occurrence probability (proportion) of term $k_i$
in the whole document collection, and the parameter $\mu$ is set to
2000 as suggested in the literature
\cite{Zhai:2001:SSM:383952.384019}.

\subsection{Recommendation evaluation}

The evaluation setup for the recommendation was designed analogously
to term-relevance prediction.  Each classifier output $f_b$ for a
block $b$ was given as input for the intent model. The resulting
intent model was used to predicted relevant words, and a ranked set of
the top-30 documents were retrieved from the whole English Wikipedia
corpus.

\subsection{Random feedback recommendation}

For a given block $b$, the recommendation was evaluated with
term-relevance input resulting from permuted relevance
judgments. Similarly to the relevance prediction, this random strategy
is also a permutation test. A small $p$-value indicates that 
the recommendation system is able to gain more relevant documents
based on the brain input than with the random input. Following the
evaluation setup of term-relevance prediction,
$k = 1000$ permutations were performed for each block.

\subsection{Performance measures}

The recommendation performance was evaluated using \textit{Cumulative
  information gain} (CG) \cite{Jarvelin:2002:CGE:582415.582418}.
The cumulative information gain is defined simply as the sum of the
relevance scores assigned by the experts for the documents that were
ranked in the top-30 documents by the retrieval system in response to
the input. Formally, 
\begin{equation}
CG = \sum\limits_{30}^{i=1} rel_i\text{,}
\end{equation}
where $rel_i$ is the relevance score of the $i$th document in the
ranked list. This measure was chosen because it allows graded relevance
assessments: some documents may be highly relevant and some documents
may be marginally relevant. The cumulative gain may be different for
different topics: some topics may have many highly relevant documents,
and some may have only a few.

{\tiny
\begin{table*}
  \resizebox{\textwidth}{!}{%
\begin{tabular}{lrrrrrrr}
  \hline
Document & \#Relevant & \#Irrelevant & \#Retrieved Documents & \#Relevant Documents & \#Irrelevant Documents & Top-30 Score & Maximum Score \\ 
  \hline
Association football & 5 & 4 & 470 & 34 & 436 & 52 & 56 \\ 
  Atom & 7 & 1 & 461 & 56 & 405 & 68 & 94 \\ 
  Automobile & 5 & 4 & 477 & 40 & 437 & 36 & 46 \\ 
  Bank & 4 & 4 & 537 & 47 & 490 & 47 & 64 \\ 
  Bicycle & 2 & 5 & 380 & 35 & 345 & 53 & 58 \\ 
  Bill Clinton & 2 & 5 & 306 & 55 & 251 & 48 & 73 \\ 
  Brain & 5 & 0 & 257 & 46 & 211 & 47 & 63 \\ 
  Cat & 7 & 4 & 545 & 55 & 490 & 66 & 91 \\ 
  Communism & 3 & 4 & 428 & 43 & 385 & 47 & 60 \\ 
  Euro & 2 & 3 & 288 & 41 & 247 & 63 & 74 \\ 
  India & 6 & 0 & 468 & 120 & 348 & 90 & 244 \\ 
  Learning & 4 & 5 & 497 & 81 & 416 & 70 & 125 \\ 
  Machine Learning & 6 & 2 & 491 & 51 & 440 & 89 & 118 \\ 
  Michael Jackson & 3 & 5 & 517 & 54 & 463 & 90 & 147 \\ 
  Money & 4 & 5 & 478 & 123 & 355 & 90 & 249 \\ 
  Ocean & 5 & 5 & 426 & 79 & 347 & 90 & 167 \\ 
  Painting & 3 & 8 & 617 & 51 & 566 & 90 & 136 \\ 
  Plato & 3 & 3 & 337 & 94 & 243 & 90 & 185 \\ 
  Politics & 5 & 6 & 588 & 172 & 416 & 90 & 337 \\ 
  Rome & 3 & 5 & 474 & 62 & 412 & 90 & 150 \\ 
  Savanna & 1 & 6 & 471 & 41 & 430 & 47 & 58 \\ 
  Schizophrenia & 6 & 3 & 484 & 51 & 433 & 69 & 90 \\ 
  School & 2 & 6 & 414 & 30 & 384 & 51 & 51 \\ 
  Society & 5 & 6 & 688 & 79 & 609 & 53 & 102 \\ 
  Star & 4 & 5 & 524 & 98 & 426 & 64 & 132 \\ 
  Telephone & 2 & 3 & 354 & 44 & 310 & 60 & 74 \\ 
  Time & 6 & 2 & 525 & 56 & 469 & 59 & 85 \\ 
  Volcano & 4 & 3 & 442 & 76 & 366 & 60 & 106 \\ 
  Wife & 2 & 5 & 450 & 66 & 384 & 49 & 85 \\ 
  Wine & 4 & 3 & 577 & 89 & 488 & 90 & 183 \\ 
  \textit{Total} & 120 & 120 & 13971 & 1969 & 12002 & 2008 & 3503 \\ 
   \hline
\end{tabular}

  }
  \caption{Description of the 30 documents used in the experiment.
    The first two columns show how often the document was presented to
    the users and how often it then was chosen as relevant or
    irrelevant.
    The third column shows the number of retrieved documents for a
    given document pooled over all experiments. The fourth and fifth
    columns show how many of the retrieved documents were judged by
    the experts to be relevant or irrelevant given the topic.
    The sixth column shows the sum of the relevance scores of the
    top-30 documents. The seventh column shows the sum of all
    relevance scores.\label{tab:document-stats}}
\end{table*}
}

\begin{table*}
  \resizebox{\textwidth}{!}{%
\begin{tabular}{lrrrrrrr}
  \hline
Participant & \#Recorded Channels & \# Accepted Channels & \#Blocks & \#Recorded Epochs & \#Accepted Epochs & \#Relevant Epochs & \#Irrelevant Epochs \\ 
  \hline
TRPB101 & 32 & 26 & 8 & 1941 & 1376 & 153 & 1223 \\ 
  TRPB102 & 32 & 26 & 8 & 1961 & 1659 & 193 & 1466 \\ 
  TRPB103 & 32 & 11 & 8 & 1936 & 1521 & 242 & 1279 \\ 
  TRPB105 & 32 & 30 & 8 & 1986 & 1521 & 198 & 1323 \\ 
  TRPB106 & 32 & 29 & 8 & 1959 & 1486 & 215 & 1271 \\ 
  TRPB107 & 32 & 20 & 8 & 1960 & 1566 & 245 & 1321 \\ 
  TRPB109 & 32 & 30 & 8 & 1869 & 1622 & 315 & 1307 \\ 
  TRPB110 & 32 & 20 & 8 & 1958 & 1021 & 103 & 918 \\ 
  TRPB111 & 32 & 31 & 8 & 1818 & 1045 & 170 & 875 \\ 
  TRPB112 & 32 & 30 & 8 & 2026 & 1588 & 268 & 1320 \\ 
  TRPB113 & 32 & 26 & 8 & 1939 & 1422 & 195 & 1227 \\ 
  TRPB114 & 32 & 26 & 8 & 1944 & 1226 & 204 & 1022 \\ 
  TRPB115 & 32 & 30 & 8 & 1896 & 1441 & 211 & 1230 \\ 
  TRPB116 & 32 & 28 & 8 & 1981 & 1662 & 242 & 1420 \\ 
  TRPB117 & 32 & 16 & 8 & 1906 & 1364 & 326 & 1038 \\ 
   \hline
\end{tabular}

  }
  \caption{Description of the EEG recordings.
    The first two columns show the number of recorded and the number
    of accepted channels after cleaning per participant.
    The third column shows the number or blocks recorded for each
    participant.
    The fourth and fith columns show the number of recorded and the
    number of accepted epochs after cleaning per participant.
    The sixth and seventh columns show the number of relevant and
    irrelvant epochs.\label{tab:eeg-stats}
  }
\end{table*}

\begin{figure*}
  \centerline{
  \includegraphics[scale=0.6]{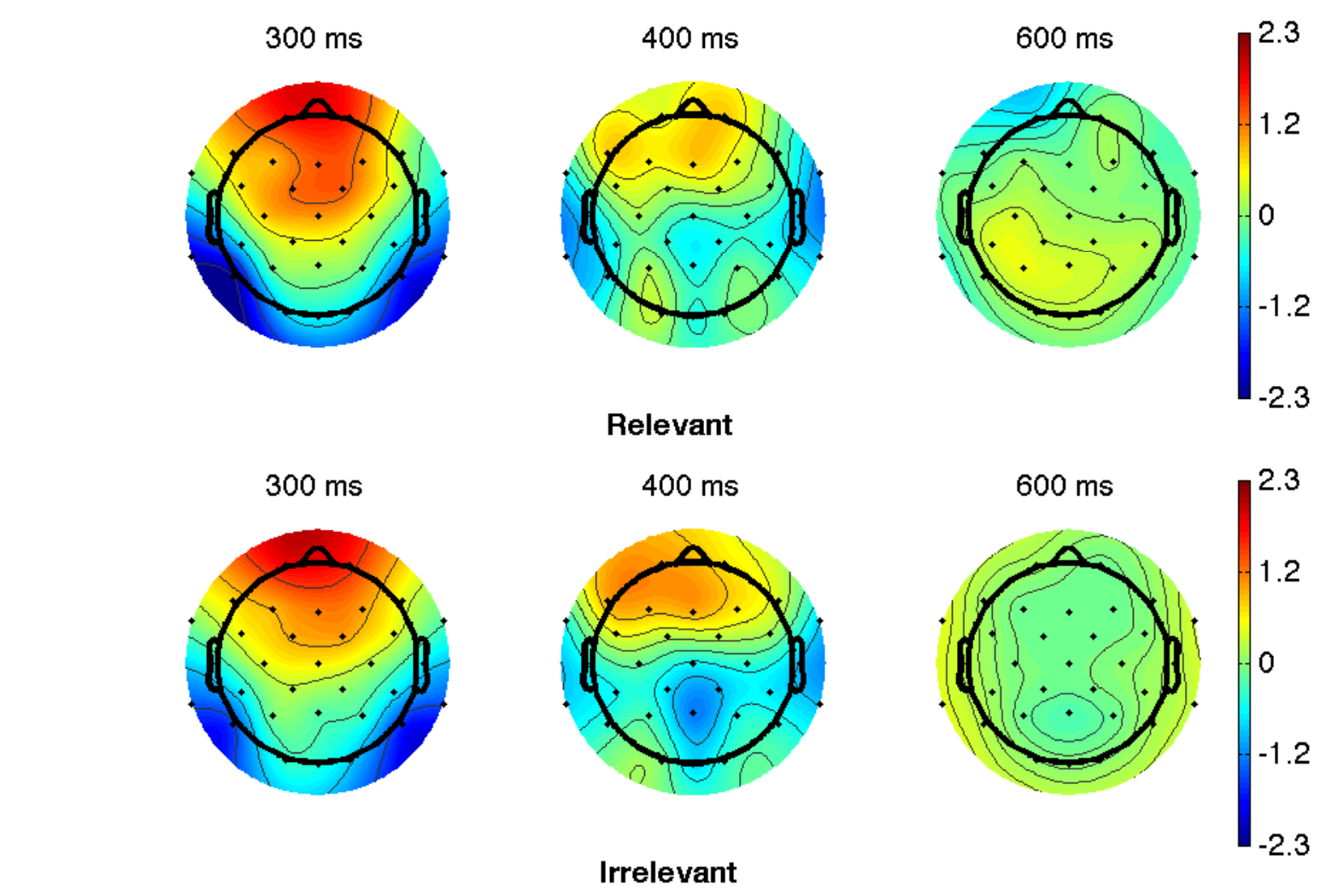}}
  \caption{Grand average-based topographic skalp plots of relevant
    words (top) and irrelevant words (bottome) for different time
    windows.\label{fig:grand-average-topo}}
\end{figure*}

\begin{figure*}
  \centerline{
    \includegraphics[scale=1]{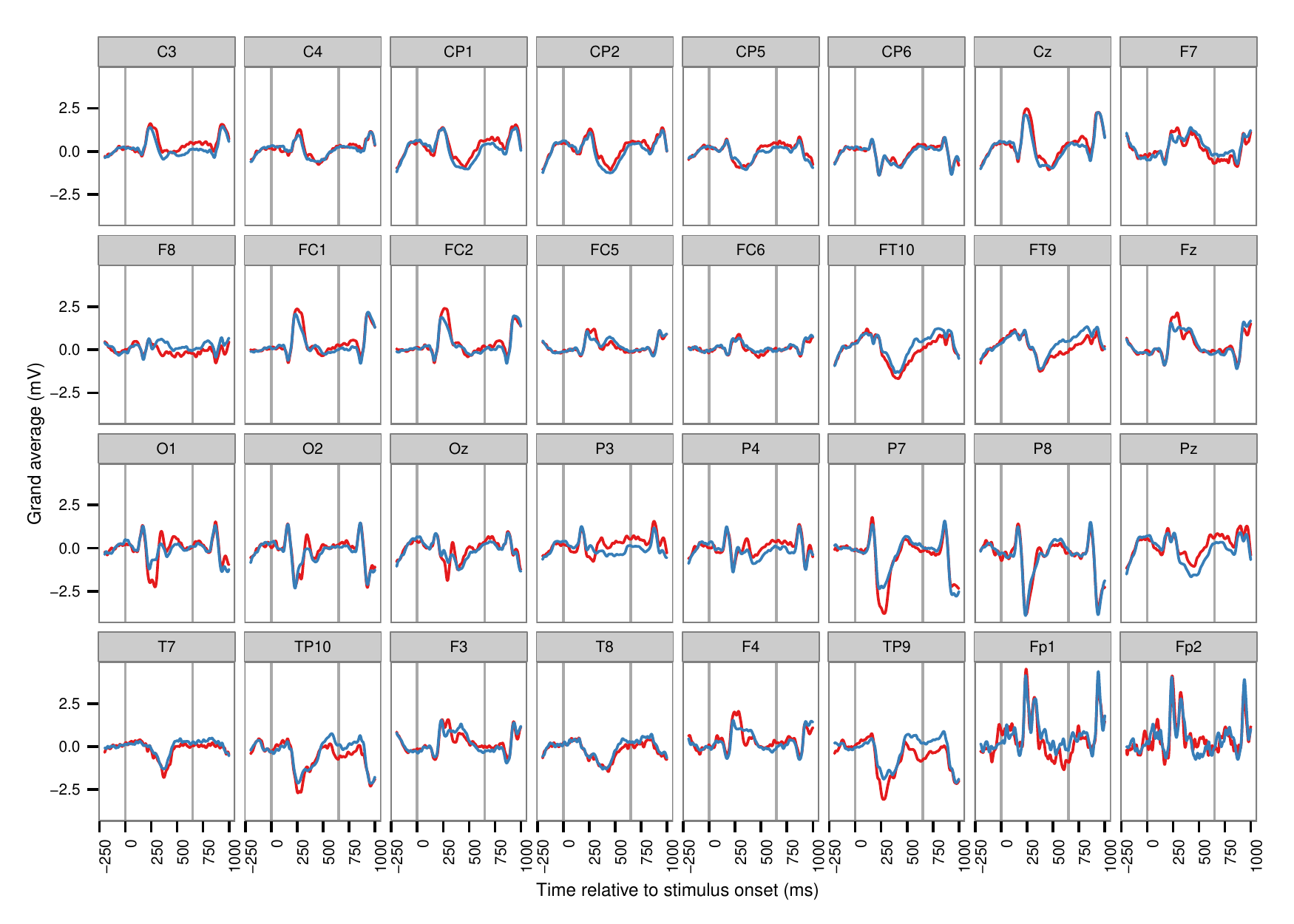}}
  \caption{Grand average event-related potential at all channels of
    relevant (red curves) and irrelevant (blue curves) terms. The gray
    vertical lines show the word onset events.\label{fig:grand-average-all}}
\end{figure*}

\begin{table}
  \centering
\begin{tabular}{ll}
  \hline
  Participant & $p$-value \\ 
  \hline
  TRPB101 & 0.0030 \\ 
  TRPB102 & 0.0010 \\ 
  TRPB103 & 0.0010 \\ 
  TRPB105 & 0.0090 \\ 
  TRPB106 & 0.0010 \\ 
  TRPB107 & 0.0010 \\ 
  TRPB109 & 0.0010 \\ 
  TRPB110 & 0.8541 \\ 
  TRPB111 & 0.0010 \\ 
  TRPB112 & 0.0010 \\ 
  TRPB113 & 0.0040 \\ 
  TRPB114 & 0.0010 \\ 
  TRPB115 & 0.0050 \\ 
  TRPB116 & 0.0010 \\ 
  TRPB117 & 0.1439 \\ 
   \hline
\end{tabular}
  \caption{Test statistics for the tests results shown in
    Figure~5. For each participant a permutation test with
    1000~iterations was executed. In each iteration, the relevance
    judgments were permutated. The $p$-value is then based on the
    number of times the randomized classification is better than the
    brain feedback-based classification with respect to the AUC
    values.\label{tab:classif-pvals}}
\end{table}

\begin{table}
\begin{tabular}{lll}
  \hline
  Participant & W & $p$-value \\ 
  \hline
  TRPB101 & 112811.5000 & 0.0009 \\ 
  TRPB102 & 111581.0000 & 0.9573 \\ 
  TRPB103 & 99082.5000 & 0.0001 \\ 
  TRPB105 & 104242.0000 & 0.4398 \\ 
  TRPB106 & 119899.0000 & 0.0000 \\ 
  TRPB107 & 94403.5000 & 0.9789 \\ 
  TRPB109 & 138593.5000 & 0.0000 \\ 
  TRPB111 & 136092.5000 & 0.0000 \\ 
  TRPB112 & 135740.0000 & 0.0000 \\ 
  TRPB113 & 127788.5000 & 0.0052 \\ 
  TRPB114 & 166286.0000 & 0.0000 \\ 
  TRPB115 & 110133.5000 & 0.0031 \\ 
  TRPB116 & 165508.0000 & 0.0000 \\ 
   \hline
\end{tabular}
  \caption{Test statistics for the tests results shown in
    Figure~7. For each participant a two-sided Wilcoxon test was
    executed between the brain feedback-based retrieved document
    scores and the random feedback-retrieved document
    scores.\label{tab:retriev-pvals}}
\end{table}

\begin{figure*}
  \centerline{
    \includegraphics[scale=1]{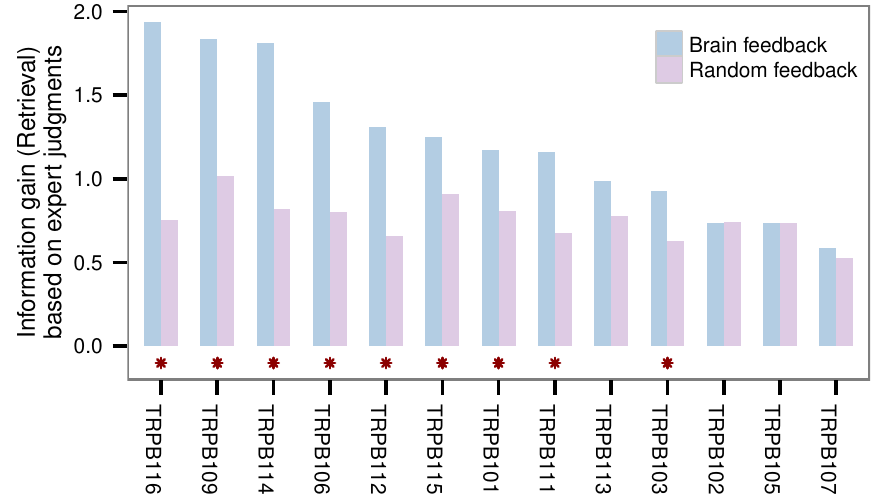} \quad %
    \includegraphics[scale=1]{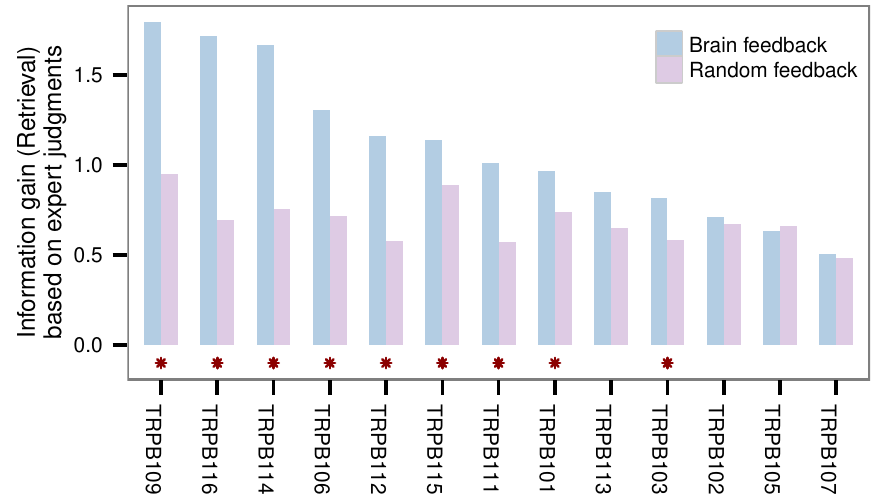}
  }
  \caption{Average cumulative information gain (on a scale 0-3) based
  on the Top-10 (left) and the Top~20 (right) retrieved documents for
  the participant. Asterisks indicate a significantly better pooled
  information gain based on brain feedback than randomized feedback
  retrieval on 1000 iterations.\label{fig:retriev-cumgain-top}}
\end{figure*}

\clearpage


\end{document}